\documentclass[aps, pra, twocolumn, preprintnumbers, superscriptaddress, showpacs, amssymb]{revtex4-1}

\usepackage{graphicx}
\usepackage{dcolumn}
\usepackage{bm}

\begin{document}

\preprint{Ver. 3p2}¡¡

\title{Upper critical field and quantum oscillations in tetragonal superconducting FeS}


\author{Taichi Terashima}
\author{Naoki Kikugawa}
\affiliation{National Institute for Materials Science, Tsukuba, Ibaraki 305-0003, Japan}
\author{Hai Lin}
\author{Xiyu Zhu}
\author{Hai-Hu Wen}
\affiliation{Center for Superconducting Physics and Materials, National Laboratory of Solid State Microstructures and Department of Physics, National Center of Microstructures and Quantum Manipulation, Nanjing University, Nanjing 210093, China}
\author{Takuya Nomoto}
\affiliation{Department of Physics, Kyoto University, Kyoto, 606-8502, Japan}
\author{Katsuhiro Suzuki}
\affiliation{Research Organization of Science and Technology, Ritsumeikan University, Kusatsu, Shiga 525-8577, Japan}
\author{Hiroaki Ikeda}
\affiliation{Department of Physics, Ritsumeikan University, Kusatsu, Shiga 525-8577, Japan}
\author{Shinya Uji}
\affiliation{National Institute for Materials Science, Tsukuba, Ibaraki 305-0003, Japan}


\date{\today}
\begin{abstract}
The magnetoresistance and magnetic torque of FeS are measured in magnetic fields $B$ of up to 18 T down to a temperature of 0.03 K.
The superconducting transition temperature is found to be $T_c$ = 4.1 K, and the anisotropy ratio of the upper critical field $B_{c2}$ at $T_c$ is estimated from the initial slopes to be $\Gamma(T_c)$ = 6.9.
$B_{c2}(0)$ is estimated to be 2.2 and 0.36 T for $B \parallel ab$ and $c$, respectively.
Quantum oscillations are observed in both the resistance and torque.
Two frequencies $F$ = 0.15 and 0.20 kT are resolved and assigned to a quasi-two-dimensional Fermi surface cylinder.
The carrier density and Sommerfeld coefficient associated with this cylinder are estimated to be 5.8 $\times$ 10$^{-3}$ carriers/Fe and 0.48 mJ/(K$^2$mol), respectively.
Other Fermi surface pockets still remain to be found.
Band-structure calculations are performed and compared to the experimental results.
\end{abstract}


\maketitle



\newcommand{\ud}{\mathrm{d}}
\def\degree{\kern-.2em\r{}\kern-.3em}

\section{Introduction}

Since Kamihara \textit{et al.} \cite{Kamihara08JACS} discovered superconductivity in LaFeAs(O$_{1-x}$F$_x$) at $T_c$ = 26 K, layered iron pnictides and chalcogenides have been studied extensively.
Very recently, an interesting new member has joined this group: tetragonal FeS, which has the same PbO-type structure as FeSe \cite{Lai15JACS}.
Instead of previous synthesis routes, which failed, Lai \textit{et al.} used a hydrothermal method and obtained highly stoichiometric tetragonal FeS crystals showing superconductivity below $T_c \approx$ 5 K.
This has aroused considerable interest and initiated research into the superconducting properties of FeS \cite{Pachmayr16ChemCommun, Borg16PRB, Yang16PRB, Xing16PRB, Holenstein16PRB, Lin16PRB, YangXiong16PRB}.
A nodal or highly anisotropic superconducting gap has been suggested by studies on the specific heat \cite{Xing16PRB} and with scanning tunneling microscopy \cite{YangXiong16PRB}, while a $\mu$SR study has suggested that a full-gap state coexists with low-moment disordered magnetism \cite{Holenstein16PRB}.
Large anisotropy of the upper critical field $B_{c2}$ has been reported \cite{Borg16PRB, Lin16PRB}.

While studies on FeS are important in themselves, comparing FeS and its sister compound FeSe is also particularly helpful in uncovering the origins of the peculiarities of FeSe.
FeSe ($T_c \approx$ 8 K \cite{Hsu08PNAS}) exhibits a tetragonal-to-orthorhombic transition, but no accompanying magnetic order occurs at ambient pressure \cite{McQueen09PRL} unlike typical iron pnictide parent compounds such as LaFeAsO or BaFe$_2$As$_2$ \cite{Rotter08PRL, Sasmal08PRL}.
Applying pressure not only enhances $T_c$ remarkably to 37 K (onset) \cite{Mizuguchi08APL, Medvedev09Nmat} but also induces an antiferromagnetic order \cite{Bendele10PRL, Bendele12PRB, Terashima15JPSJ}.
The electronic structure is markedly different from that predicted by band-structure calculations: the Fermi surface (FS) is anomalously small \cite{Maletz14PRB, Terashima14PRB, Audouard15EPL, Watson15PRB, Watson15PRL}, and a large split of the Fe 3$d_{xz}$ and $d_{yz}$ bands has been reported \cite{Tan13NatMat, Shimojima14PRB, Nakayama14PRL}.
The smallness of the Fermi energy $E_F$ relative to $T_c$ suggests that the physics of the Bardeen--Cooper--Schrieffer to Bose--Einstein condensation (BCS--BEC) crossover may be relevant to the superconducting properties of FeSe \cite{Kasahara14PNAS, Terashima16PRB, Terashima16PRB2}.

In this paper, we report on the magnetoresistance and magnetic torque measurements of FeS crystals.
The upper critical field $B_{c2}$ is determined from the magnetoresistance data for magnetic fields $B$ parallel to the $c$ axis and $ab$ plane down to a temperature of 0.03 K.
Quantum oscillations are observed in both the magnetoresistance and magnetic torque at high fields.
The results are discussed in terms of band-structure calculations.

\section{Experiments}

Tetragonal FeS single crystals were prepared by a hydrothermal method, as described in Ref. \cite{Lin16PRB}.
For four-contact in-plane resistance measurements, electrical contacts were first attached to four samples with silver paste.
However, the contacts changed color to black, and only one could be measured down to low temperatures.
The contacts appeared to have been degraded by the formation of AgS.
Then, carbon paste was used for the other three samples, two of which could be measured at low temperatures.
Typical sample dimensions of the resistance samples were roughly 0.7 $\times$ 0.2 $\times$ 0.02 mm$^3$.
The resistance ratio between room temperature and 4.5 K was 14--25.
The magnetic torque was measured on four samples with typical dimensions of 0.2 $\times$ 0.2 $\times$ 0.02 mm$^3$ by using piezoresistive microcantilevers \cite{Ohmichi02RSI}.
A dilution refrigerator and superconducting magnet were used to generate low temperatures down to $T$ = 0.03 K and high magnetic fields up to $B$ = 17.8 T.
The magnetic field direction $\theta$ was measured from the $c$ axis.
Relativistic electronic structure calculations were performed by using the WIEN2K code \cite{WIEN2K} with the experimental lattice parameters \cite{Lai15JACS, *[{We used the PBE-GGA exchange-correlation functional given in }] [{ and set $R_{MT}K_{max}=7.0$ and $8 \times 8 \times 6$ $k$-mesh in the Brillouin zone.}] Perdew96PRL}


\section{Results and discussion}

\begin{figure}
\includegraphics[width=8.6cm]{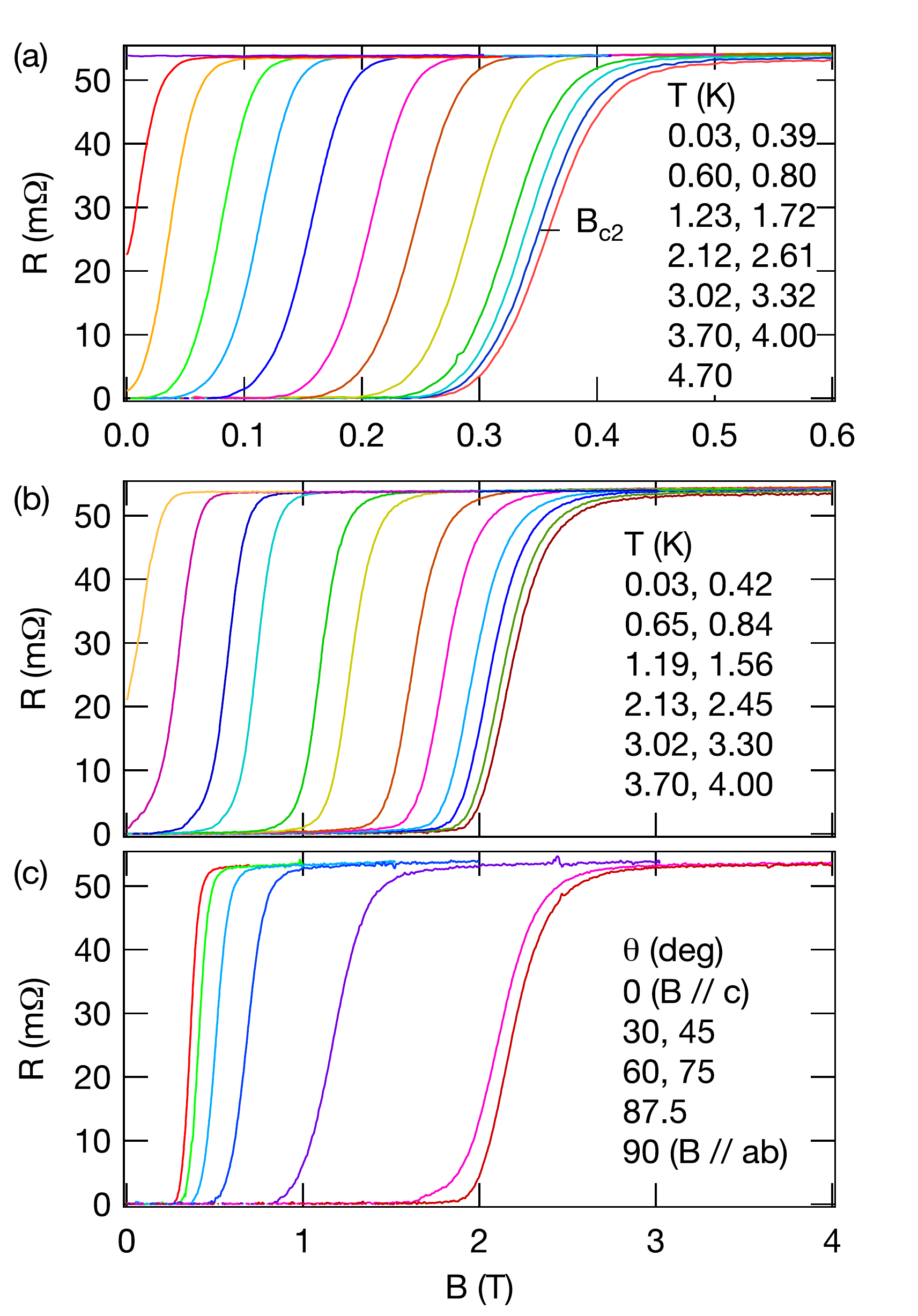}
\caption{\label{RvsB}(Color online). In-plane resistance $R$ in FeS as a function of the magnetic field $B$ parallel to the (a) $c$ axis and (b) $ab$ plane for selected temperatures. $B_{c2}$ is defined with a midpoint criterion, as indicated in (a). (c) $R$ vs $B$ for selected field directions. $T \leqslant 0.04$ K.}
\end{figure}

\begin{figure}
\includegraphics[width=8.6cm]{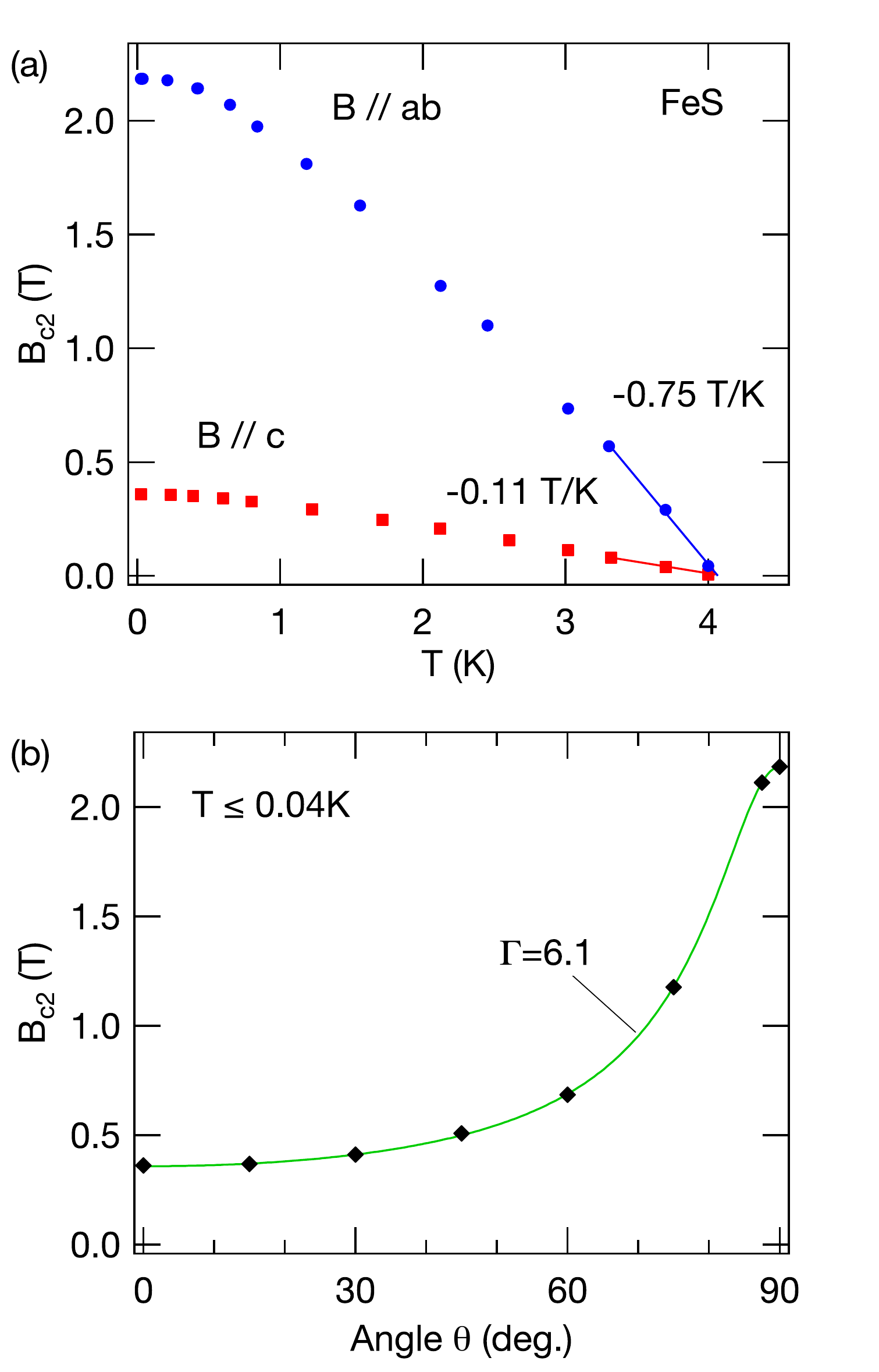}
\caption{\label{Bc2}(Color online). (a) Temperature dependence of $B_{c2}$ for $B \parallel ab$ and $B \parallel c$. (b) Angle dependence of $B_{c2}$. The solid curve is calculated with an anisotropic Ginzburg--Landau model.}
\end{figure}

Figure~\ref{RvsB} shows the resistive transition curves for (a) $B \parallel c$ and (b) $B \parallel ab$ at selected temperatures and (c) selected field directions at $T \leqslant 0.04$ K.
We define $B_{c2}$ by a midpoint criterion, as indicated in (a).
Because the resistive transition curves do not broaden appreciably in applied fields, this definition is expected to give sufficiently accurate estimates of $B_{c2}$.
Figure~\ref{Bc2}(a) shows $B_{c2}$ as a function of the temperature for $B \parallel ab$ and $B \parallel c$.
The initial slopes $\mathrm{d}B_{c2}/\mathrm{d}T|_{T_c}$ for $B \parallel ab$ and $B \parallel c$ are estimated from linear fitting to data points for $T \geqslant 3.3$ K (solid lines) to be -0.75 and -0.11 T/K, respectively; their ratio gives the anisotropy ratio $\Gamma(T_c)$ = 6.9.
The coherence length $\xi$ is estimated to be 33 and 4.7 nm for $\parallel ab$ and $\parallel c$, respectively.
The intercepts of the fitted lines give $T_c$ = 4.1 K.
The upper critical field as $T \rightarrow 0$ is $B_{c2}(0)$ = 2.2 and 0.36 T for $B \parallel ab$ and $B \parallel c$, respectively, to give $\Gamma(0)$ = 6.1.
In the case of FeSe, $B_{c2}$ for $B \parallel ab$ is anomalously enhanced at the lowest temperatures below $\sim$1 K \cite{Terashima14PRB}, while no such enhancement is observed in FeS.
An anisotropic Ginzburg--Landau model with $\Gamma(0)$ = 6.1 can be used to explain the angle dependence of the upper critical field measured at $T \leqslant 0.04$ K, as shown in Fig.~\ref{Bc2}(b).
These superconducting parameters are in good agreement with \cite{Lin16PRB}.
On the other hand, the anisotropy ratio of $\Gamma \sim$ 10 reported by Borg et al.~\cite{Borg16PRB} is significantly larger than our value.
This may be related to the fact that $T_c$ of their resistivity sample \cite{Borg16PRB} is rather lower than our $T_c$; they reported that $T_c^{\mathrm{onset}} = 3.5$ K and $T_c^{\mathrm{zero}} = 2.4$ K.

It is interesting to compare the superconducting parameters of FeS and FeSe.
The initial slopes in FeSe are much larger \cite{Terashima14PRB}: $\mathrm{d}B_{c2}/\mathrm{d}T|_{T_c}$ = -6.9 and -1.6 T/K for $B \parallel ab$ and $B \parallel c$, respectively, which are factors of 9.1 and 15 larger than the corresponding values in FeS.
Within single-band BCS theory, $\mathrm{d}B_{c2}/\mathrm{d}T|_{T_c} \sim \gamma^2 T_c/S^2$, where $S$ is the surface area of the FS \cite{Decroux82Book}.
To evaluate the right-hand side of the relation, we use the Sommerfeld coefficient of $\gamma$ = 5.73 mJ/(mol K$^2$) \cite{Lin11PRB} and $T_c$ = 9.1 K for FeSe \cite{Terashima14PRB}.
For FeS, $\gamma$ = 3.8 mJ/(mol K$^2$) \cite{Xing16PRB}.
Because $T_c$ for FeSe was determined with a zero-resistance criterion in \cite{Terashima14PRB}, we use a similarly determined $T_c$ of 3.9 K for FeS for the comparison.
The ratio of $\gamma^2 T_c/S^2$ between FeSe and FeS is evaluated to be 5.2$(S_{FeS}/S_{FeSe})^2$.
The experimental ratio of 9.1 or 15 suggests $(S_{FeS}/S_{FeSe})$ = 1.3 or 1.7, which may suggest that FeS has a larger FS and hence a larger carrier density.
Second, the anisotropy ratio $\Gamma(T_c)$ = 6.9 is larger than the ratio of 4.3 in FeSe.
Because $\Gamma^2 = m_c/m_{ab}$, where $m_{c (ab)}$ is the effective mass along the $c$ ($ab$) direction, this suggests that FeS has a more two-dimensional electronic structure, i.e., smaller dispersion along the $c$ axis.

\begin{figure}
\includegraphics[width=8.6cm]{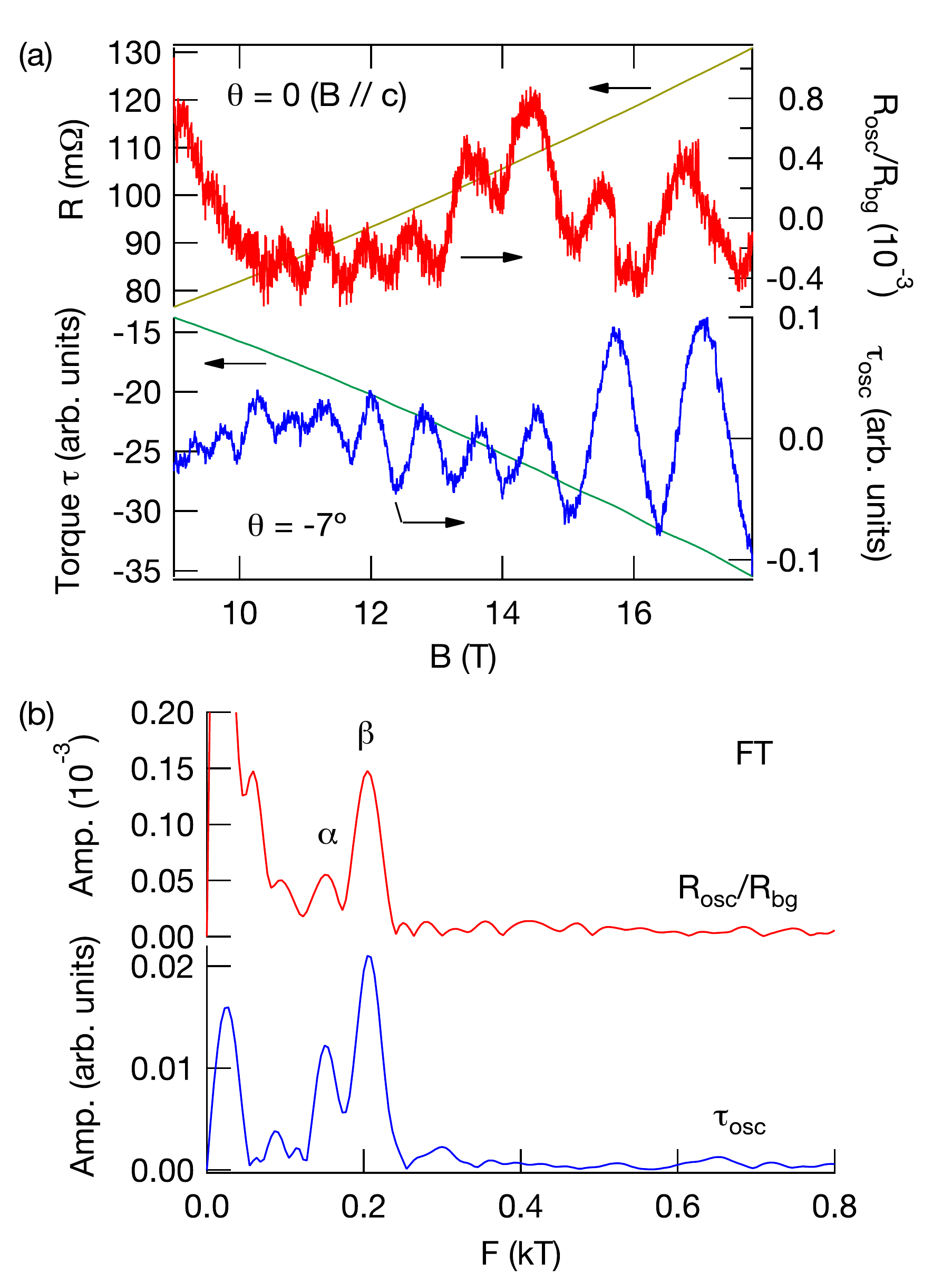}
\caption{\label{Sig_FT}(Color online). (a) Magnetoresistance in FeS for $B \parallel c$ at $T$ = 0.11 K (upper curve) and magnetic torque for $\theta = -7^{\circ}$ at $T$ = 0.07 K (lower curve) as a function of the magnetic field. Their oscillatory parts are also shown (right axis). (b) Fourier transform amplitudes of the oscillations in $1/B$.}
\end{figure}

\begin{figure}
\includegraphics[width=8.6cm]{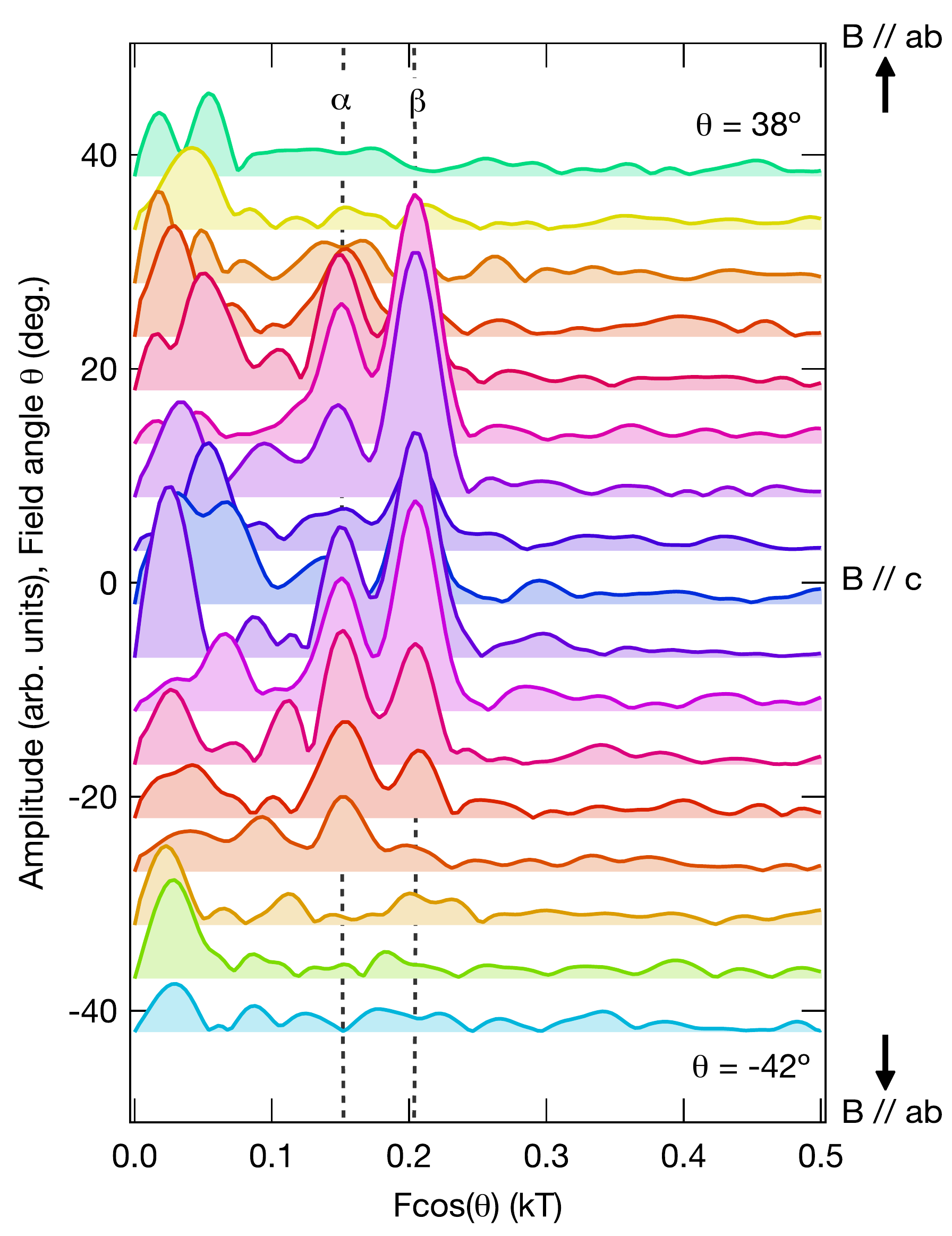}
\caption{\label{FT_ang}(Color online). Fourier transform amplitudes of torque dHvA oscillations in FeS for different field directions. Note that the horizontal axis is $F\cos \theta$. Spectra are shifted vertically so that the baseline of a spectrum for the angle $\theta$ is placed at $\theta$. The dotted lines indicate the $\theta$-variation of $F\cos \theta$ calculated with the Yamaji model of a quasi-two-dimensional FS cylinder (see text).}
\end{figure}

Owing to the high quality of the crystals, the magnetoresistance exhibits Shubnikov--de Haas (SdH) oscillations at high fields.
Figure~\ref{Sig_FT}(a) shows an $R(B)$ curve for $B \parallel c$ (upper curve).
After subtraction of a smooth background modeled by a second-order polynomial, clear oscillations appear.
The corresponding Fourier spectrum shows two peaks marked by $\alpha$ and $\beta$ [Fig.~\ref{Sig_FT}(b), upper curve].

To further study the FS in FeS, we also measured the magnetic torque.
The lower part of Fig.~\ref{Sig_FT}(a) shows the magnetic torque at $\theta$ = -7$^{\circ}$ and its oscillatory part, i.e., de Haas--van Alphen (dHvA) oscillations, as a function of $B$.
The smooth background is modeled with a third-order polynomial.
The corresponding Fourier spectrum [Fig.~\ref{Sig_FT}(b), lower curve] shows $\alpha$ and $\beta$ frequencies consistent with the SdH data.
Figure~\ref{FT_ang} shows the angular variation of the Fourier spectra.
Note that the horizontal axis is $F\cos\theta$.
The symmetric appearance of the frequency peaks with respect to $\theta$ = 0 ($B \parallel c$) and the suppressed oscillation amplitudes near $\theta$ = 0 conform to the crystal symmetry of FeS.

All of the three resistance and four torque samples exhibit quantum oscillations with the $\alpha$ and $\beta$ frequencies, which demonstrates that the two frequencies are intrinsic to tetragonal FeS.
The effective masses associated with the two frequencies are estimated from the temperature dependence of the oscillation amplitudes using the conventional method \cite{Shoenberg84} (for more details, see the Appendix).
Averaged over four samples with larger oscillation amplitudes, the frequencies and effective masses for $B \parallel c$ are estimated as follows:
$F_{\alpha}$ = 153(2) T, $m^*_{\alpha}/m_e$ = 0.62(3), $F_{\beta}$ = 203.5(2) T, and $m^*_{\beta}/m_e$ = 0.83(1), where $m_e$ is the free-electron mass.
The frequencies correspond to the orbit areas occupying only 0.50\% and 0.67\% of the Brillouin zone.
The effective Fermi energy $E_F$ can be estimated from experimental values of $F$ and $m^*$ by using the following formulas:
$E_F= \hbar^2 k_F^2/(2m^*)$, $A=\pi k_F^2$, and $F=\hbar A/(2\pi e)$, where $A$ is the orbit area in the $k$ space and we assume circular orbits.
This estimation gives $E_F$ = 28 meV for both orbits.
The ratio $k_BT_c/E_F$ is estimated to be $\sim$0.01, which is much smaller than the values found in FeSe: 0.04--0.22 \cite{Terashima14PRB, Terashima16PRB}.
This suggests that, unlike FeSe, FeS is not close to the BCS--BEC crossover (for single-band superconductors, $k_BT_c/E_F$ = 0.2 would indicate the crossover \cite{Randeria14ARCMP}).
The electron mean free path $l$ can be estimated only very roughly because of the small number of observed oscillation periods: $l \approx$ 40--80 nm for the $\beta$ orbit.
This is comparable to or slightly larger than the in-plane coherence length.

We return to Fig.~\ref{FT_ang}.
We assign the two frequencies $\alpha$ and $\beta$ to the minimum and maximum cross-sections of a quasi-two-dimensional FS cylinder.
For a purely two-dimensional FS cylinder, there is a single quantum oscillation frequency $F$, and $F\cos\theta$ (the horizontal axis of Fig.~\ref{FT_ang}) remains constant as $\theta$ is varied.
However, there is a $c$-axis energy dispersion in real materials that modulates the cylinder, and two frequencies corresponding to the maximum and minimum cross-sections of the modulated cylinder will appear.
Yamaji calculated the angle dependence of $F\cos\theta$ for the two frequencies by assuming a cosine energy dispersion along the $c$ axis \cite{Yamaji89JPSJ}.
The dotted lines in Fig.~\ref{FT_ang} show the angle dependence of $F\cos\theta$ for the $\alpha$ and $\beta$ frequencies expected from the Yamaji model.
The observed frequency peaks are consistent with the calculated lines, which supports our assignment.
The two-dimensionality of an FS cylinder may be judged from $\Delta F/ F_{av}$, where $\Delta F$ and $F_{av}$ are the difference and average, respectively, of the minimum and maximum frequencies.
This parameter is 0.28 for the present FS cylinder in FeS, while it is larger than 1 for experimentally observed FS cylinders in FeSe \cite{Terashima14PRB}.
This suggests that FeS is more two-dimensional in the electronic structure, which is consistent with larger $B_{c2}$ anisotropy.
The carrier density and Sommerfeld coefficient associated with the observed FS cylinder are estimated to be 5.8 $\times$ 10$^{-3}$ carriers/Fe and 0.48 mJ/(K$^2$mol).
Because the experimental Sommerfeld coefficient is $\gamma_{exp}$ = 3.8 mJ/(K$^2$mol) \cite{Xing16PRB}, large parts of the FS still remain to be observed in future measurements.

\begin{figure}
\includegraphics[width=8.6cm]{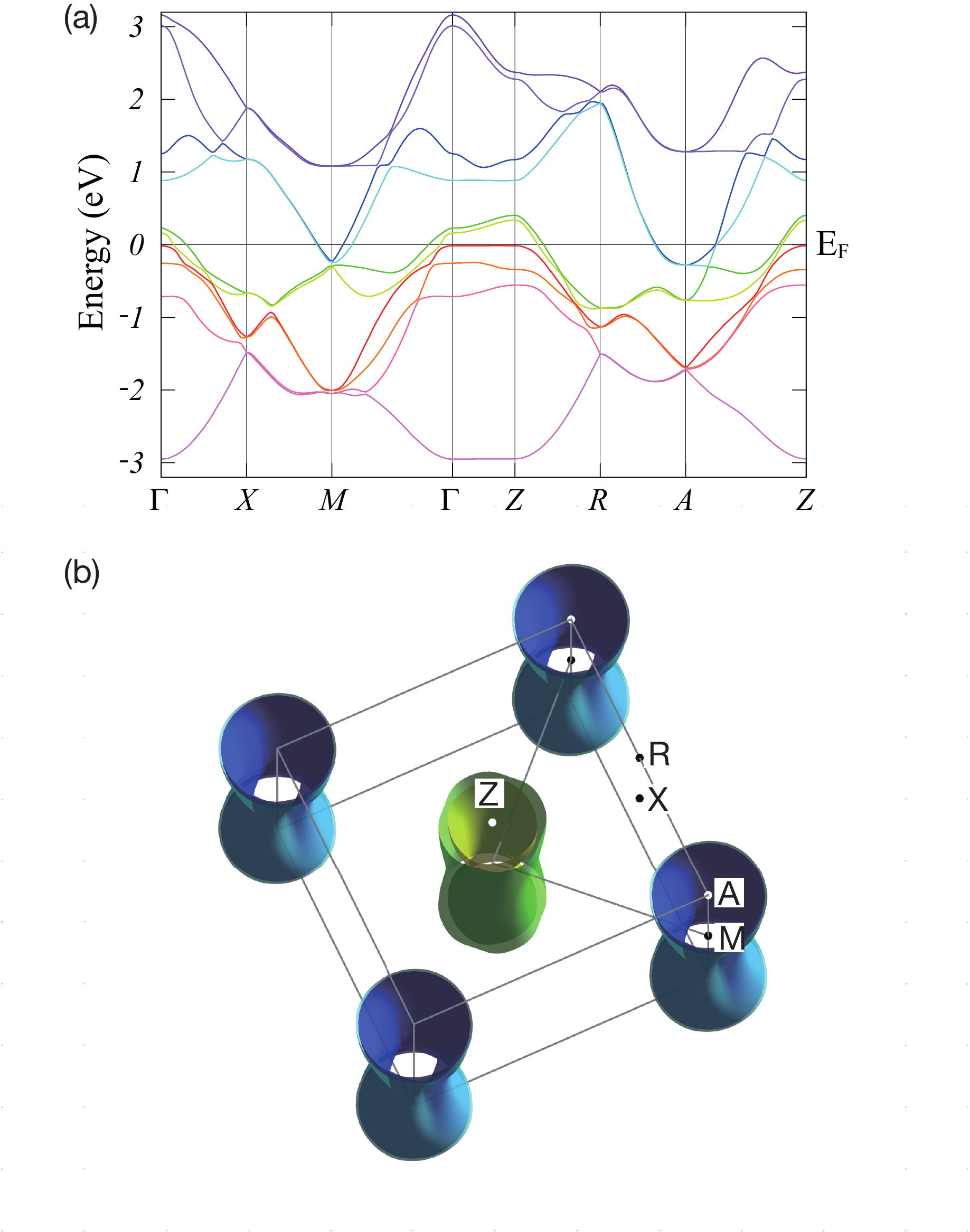}
\caption{\label{Band}(Color online). Calculated electronic band structure (a) and Fermi surface (b) in FeS.}
\end{figure}

Figure~\ref{Band}(a) shows the calculated band structure.
The calculated Fermi surface (b) consists of two hole and two electron FS cylinders at the zone center and corner, respectively.
The band structure and Fermi surface are in good agreement with \cite{Yang16PRB} but not with \cite{Subedi08PRB_FeSe}.
The latter suggested an additional closed hole pocket at $\Gamma$.
The discrepancy can be attributed to the difference in the atomic position $z_{S}$ of S: while the experimental value of $z_S$ = 0.2523 \cite{Lai15JACS} was used in the present work and \cite{Yang16PRB}, the relaxed value of $z_S$ = 0.2243 was used in \cite{Subedi08PRB_FeSe}.
The carrier density and Sommerfeld coefficient are estimated to be $n_e$ = $n_h$ = 0.185 carriers/Fe and $\gamma_{band}$ = 2.4 mJ/(K$^2$mol).
The carrier density is slightly larger than that in FeSe (0.17 carriers/Fe) \cite{Terashima14PRB}.
The latter gives the mass enhancement of $\gamma_{exp}/\gamma_{band} = 1+ \lambda$ = 1.6.
Clearly, the calculated FS cylinders are much larger than the experimentally observed one: the calculated quantum oscillation frequencies are in a range between $F$ = 0.5 and $\sim$3 kT.
This may indicate the FS shrinking \cite{Ortenzi09PRL}; the experimentally observed FS in FeSe and iron pnictides is smaller than predicted by the band-structure calculations \cite{Terashima14PRB, Coldea08PRL, Analytis09PRL, Shishido10PRL, Analytis10PRL, Terashima11PRL, Putzke12PRL, Yi09PRB}.
Alternatively, the observed FS cylinder may be attributed to the third hole band, which is nearly flat and sits just below $E_F$ along the $\Gamma Z$ section.
If this band is slightly raised, this will produce a fairly two-dimensional FS cylinder.
Indeed, the corresponding band sits above $E_F$ in the calculated band structure of FeSe \cite{Terashima14PRB}.

\section{Summary}

We have measured the magnetoresistance of FeS down to 0.03 K and determined the upper critical field:
$B_{c2}(0)$ = 2.2 and 0.36 T for $B \parallel ab$ and $c$, respectively.
The anisotropy ratio at $T_c$ is $\Gamma(T_c)$ = 6.9, which is consistent with \cite{Lin16PRB}.
We have observed quantum oscillations in both the magnetoresistance and magnetic torque.
Two frequencies $F$ = 0.15 and 0.20 kT are resolved and attributed to a quasi-two-dimensional FS cylinder.
The associated carrier density and Sommerfeld coefficient are estimated to be 5.8 $\times$ 10$^{-3}$ carriers/Fe and 0.48 mJ/(K$^2$mol).
Band-structure calculations predict FS cylinders that are much larger than the observed one.
A very important future task is finding remaining FS cylinders by using higher magnetic fields to judge how successful the calculations are.

\begin{acknowledgments}
This work was supported by JSPS KAKENHI Grant Numbers JP26400373, 15H05745, 15H02014, 16H01081, 16H04021, and 15J01476.
The work in Nanjing University is supported by National Natural Science Foundation of China (NSFC) with the projects A0402/11534005, A0402/11190023 and by the Ministry of Science and Technology of China (Grant Nos. 2016YFA0300404, 2012CB821403).
\end{acknowledgments}

\appendix

\section{Effective mass determination}
For the effective-mass measurements, the field was applied within 10$^{\circ}$ of the $c$ axis.
The effective masses at the measurement direction $\theta$ were determined by fitting the Lifshitz--Kosevich formula to the temperature dependences of the oscillation amplitudes \cite{Shoenberg84}.
The effective masses at $\theta = 0$ $(B \parallel c)$ were estimated by assuming $\cos\theta$ dependence, i.e., $m^*(0) = m^*(\theta)\cos\theta$, as indicated in Table~\ref{Tab}.

\begin{table}
\caption{\label{Tab} Frequencies and effective masses for $B\parallel c$. SdH measurements were performed on sample 1, while torque dHvA measurements were performed on the other samples.}
\begin{ruledtabular}
\begin{tabular}{ccccc}
sample & $F_{\alpha}$ (T) & $m^*_{\alpha}/m_e$ & $F_{\beta}$ (T) & $m^*_{\beta}/m_e$\\
\hline
1 & 153(5) & 0.7(2) & 205(3) & 0.86(6)\\
2 & 156(2) & 0.67(9) & 203(1) & 0.85(5)\\
3 & 150(5) & 0.61(9) & 203(2) & 0.82(3) \\
4 & 149(2) & 0.57(8) & 204(1) & 0.82(3)\\
\end{tabular}
\end{ruledtabular}
\end{table}

%

\end{document}